\documentclass[Physsubmission, Phys]{SciPost}

\hypersetup{
    colorlinks,
    linkcolor={red!50!black},
    citecolor={blue!50!black},
    urlcolor={blue!80!black}
}

\usepackage{xspace}
\usepackage{amssymb,amscd}
\usepackage{placeins}
\usepackage{setspace}
\usepackage[bitstream-charter]{mathdesign}
\DeclareSymbolFont{usualmathcal}{OMS}{cmsy}{m}{n}
\DeclareSymbolFontAlphabet{\mathcal}{usualmathcal}

\newcommand{\BROADFIGWIDTH}{0.50\textwidth}

\graphicspath{{Figures/}{}}

\include{MOM-tab}

\newcommand{\TTbMOMresultNNNLO}{0.12911\pm 0.00177 {\text( exp.)}\pm 0.0123 {\text( scale)}}

\newcommand{\eVdist}{\kern-0.06667em}
\newcommand{\GeV}{{\,\text{Ge}\eVdist\text{V\/}}}

\begin{document}
\begin{center}{\Large \textbf{
Determination of $\alpha_{S}$ beyond $NNLO$  using the event shape averages\\
}}\end{center}

\begin{center}
Adam Kardos\textsuperscript{1},
G\'abor Somogyi\textsuperscript{2} and
Andrii Verbytskyi\textsuperscript{3$\star$}
\end{center}

\begin{center}
{\bf 1} University of Debrecen, 4010 Debrecen, PO Box 105, Hungary
\\
{\bf 2} MTA-DE Particle Physics Research Group, 4010 Debrecen, PO Box 105, Hungary
\\
{\bf 3} Max-Planck-Institut f\"{u}r Physik, D-80805 Munich, Germany 
\\
* andrii.verbytskyi@mpp.mpg.de
\end{center}

\begin{center}
\today
\end{center}


\definecolor{palegray}{gray}{0.95}
\begin{center}
\colorbox{palegray}{
  \begin{tabular}{rr}
  \begin{minipage}{0.1\textwidth}
    \includegraphics[width=30mm]{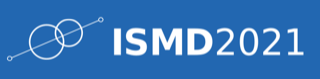}
  \end{minipage}
  &
  \begin{minipage}{0.75\textwidth}
    \begin{center}
    {\it 50th International Symposium on Multiparticle Dynamics}\\ {\it (ISMD2021)}\\
    {\it 12-16 July 2021} \\
    \doi{10.21468/SciPostPhysProc.?}\\
    \end{center}
  \end{minipage}
\end{tabular}
}
\end{center}

\section*{Abstract}
{\bf
In this proceedings we discuss a prescription to extract the QCD strong coupling constant 
at $N^{3}LO$ precision in perturbative QCD using a combination of 
 ${\cal{O}}(\alpha_{S}^{3})$ 
calculations in pQCD and estimations of the ${\cal{O}}(\alpha_{S}^{4})$ corrections from  the data. 
The method is applied to a set of  event shape  averages  measured in experiments at the LEP, PETRA, PEP and TRISTAN colliders. 
In our analysis we account for hadronization effects with models from modern 
Monte Carlo event generators and analytic hadronization models. 
We conclude that  the precision of the $\alpha_{S}$ 
extraction cannot be improved significantly only with pQCD predictions of  higher orders, 
and further progress in these studies requires a significant advances in the  studies and modeling  of hadronization process.
}

\vspace{10pt}
\noindent\rule{\textwidth}{1pt}
\tableofcontents\thispagestyle{fancy}
\noindent\rule{\textwidth}{1pt}
\vspace{-5pt}
\section{Introduction}
\label{sec:intro}
The process of hadroproduction in $e^{+}e^{-}$ annihilations
is one of the best environments for verification theoretical predictions of Quantum Chromodynamics (QCD). 
In the past, multiple  comparisons of experimental measurements of event shape and jet observables 
to perturbative QCD (pQCD) predictions were performed.  All of these comparisons were done using the data from now retired experiments.
Due to an absence of active high-energy $e^{+}e^{-}$ experiments  new data will not be available in the next decade(s) and  the improvements 
 in QCD studies in  $e^{+}e^{-}$ collisions will depend only on the advances of the theory (and phenomenology).
  However, most QCD studies of the hadroproduction in $e^{+}e^{-}$ with the available data show relatively low impact of the experimental uncertainties 
 on the results 
  in comparison to the pQCD- and modeling-related uncertainties.   

 In this situation it is interesting if 
pQCD calculations and/or resummation techniques will be able to improve the precision of 
the results (e.g.\ $\alpha_{S}(M_{Z})$) without any new data. 
And if not, what would be the limiting factors for 
the precision of QCD studies in the future and what should be done to eliminate them? 
To answer these questions we perform an extraction of $\alpha_{S}(M_Z)$ using 
estimations of higher-order corrections.

As of 2021, the calculations for the  $e^{+}e^{-}\rightarrow Z/\gamma\rightarrow jets$ process are available 
in high precision in pQCD, i.e.\ the 
fully differential predictions for the $e^{+}e^{-}\rightarrow Z/\gamma\rightarrow 3 jets$ process are available at ${\cal O}(\alpha_{S}^{3})$
and the total cross-section $e^{+}e^{-}\rightarrow Z/\gamma\rightarrow jets$ at 
${\cal O}(\alpha_{S}^{4})$~\cite{GehrmannDeRidder:2007hr,Baikov:2012zn,DelDuca:2016csb}.
Obviously, the impact of higher order corrections on the QCD analyses (e.g.\ extraction of  $\alpha_{S}$) has a significant interest.

With a sufficient amount of data, proper selection of desired observables it is possible to go beyond the pQCD
 accuracy of predictions available from the exact calculations.
This could be done with simultaneous fits of $\alpha_{S}(M_Z)$ and the ${\cal O}(\alpha_{S}^{4})$ coefficients 
not available in the exact pQCD predictions.
The approach is obviously limited to cases with only a 
small number of coefficients of the perturbative expansion to be estimated. 
In these proceeding we describe the results of implementation of this approach using the 
averages of thrust and  $C$-parameter event shape observables. 
The argumentation for the choice of observables is given in  Ref.~\cite{this}.

\section{Theory predictions and hadronization models}
\label{sec:theory}
The experimentally measured  averages of event shape observables (i.e.\ first moments)  $O$,  $\langle O^{1} \rangle$  are normalized to the total 
hadronic cross section, and the perturbative expansion of predictions for these quantities
up to  ${\cal O}(\alpha_{S}^{4})$ in the massless QCD\footnote{The prescription on the treatment of effects related to massive $b$-quarks up to NLO given in Ref.~\cite{this}.}
reads:
$$\langle O^{1} \rangle
=\frac{\alpha_{S}(\mu_0)}{2\pi}\bar{A}^{\langle O^{1} \rangle}_0
+\left(\frac{\alpha_{S}(\mu_0)}{2\pi}\right)^2 \bar{B}^{\langle O^{1} \rangle}_0
+\left(\frac{\alpha_{S}(\mu_0)}{2\pi}\right)^3 \bar{C}^{\langle O^{1} \rangle}_0
+\left(\frac{\alpha_{S}(\mu_0)}{2\pi}\right)^4 \bar{D}^{\langle O^{1} \rangle}_0. 
$$
In the later expression  the coefficients $\bar{A}^{\langle O^{1} \rangle}_0$,  $\bar{B}^{\langle O^{1} \rangle}_0$,  
$\bar{C}^{\langle O^{1} \rangle}_0$~\cite{this} are known exactly and the coefficient  $\bar{D}^{\langle O^{1} \rangle}_0$ can be extracted from
 a simultaneous fit of multiple data points at different 
center-of-mass energies performed with four-loop running of $\alpha_{S}(\mu)$. 
In the fit procedure it is also essential to take into account the hadronization effects  and we consider two types of models to
handle this problem.
Namely, we consider the Monte Carlo event generator (MCEG) models generating 
predictions at NLO accuracy in pQCD and in addition to that we consider 
 analytic hadronization models~\cite{this} extended to ${\cal O}(\alpha_{S}^{4})$ for the first time.

\begin{figure}[htbp]\centering
\includegraphics[width=\BROADFIGWIDTH]{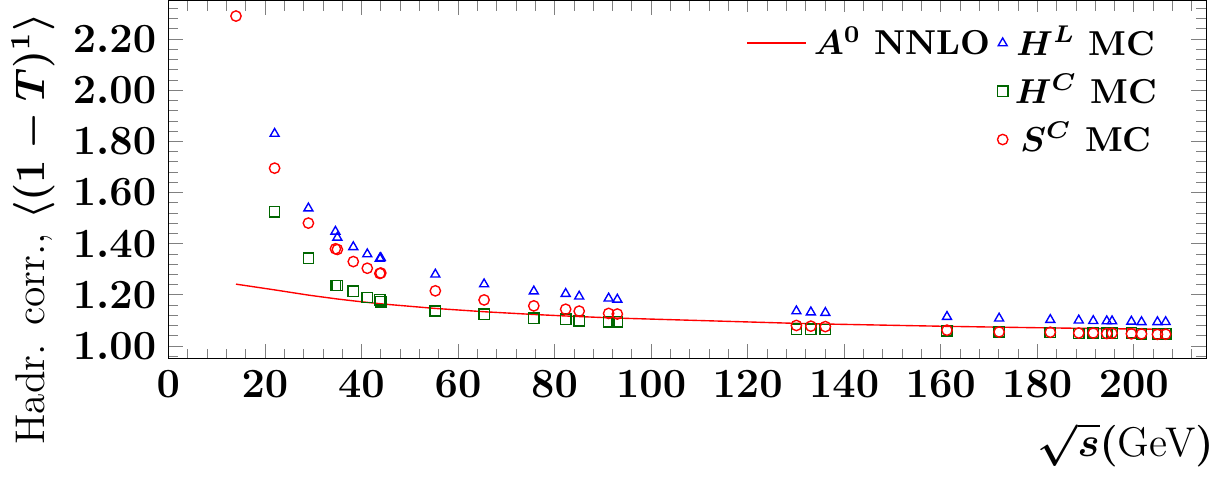}\includegraphics[width=\BROADFIGWIDTH]{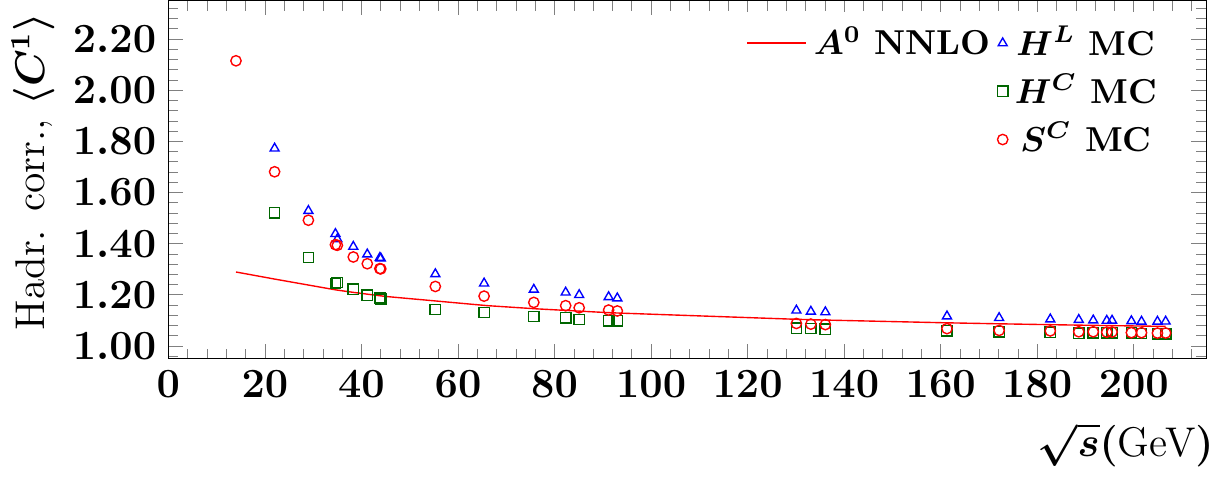}\\
\caption{Multiplicative hadronization corrections extracted from MCEGs and analytic (${\cal A}_{0}$ scheme) hadronization models~\protect\cite{this}. 
Figures from Ref.~\protect\cite{this}.}
\label{fig:had}
\end{figure}

The hadron level predictions by the MCEGs for the averages of event shape observables~\cite{this}   reasonably 
well describe the data  for a wide range of 
center-of-mass energies. 
Contrary, the corresponding quantities at the  parton level are reasonable only for $\sqrt{s}>29\GeV$.
Therefore, as the analysis was aiming also at  comparison of the used hadronization models, the fits were performed 
only to the data with $\sqrt{s}>29\GeV$.
The procedure to correct the pQCD predictions for hadronization effects with the MCEGs models consisted of applying $\sqrt{s}$-depending  factor 
 which was the ratio of values of event shape averages on the hadron and parton levels, see Fig.~\ref{fig:had}. 

The models based on the dispersive model of analytic hadronization corrections 
for event shapes~\cite{Dokshitzer:1995qm} differ in their application from the models based on the MCEG approach. The dispersive model predicts the hadron-level 
differential distributions of the event shapes can be obtained from the pQCD differential distributions with a simple shift 
$\frac{\mathrm{d}\sigma_{\mathrm{hadrons}}(O)}{\mathrm{d}O} 
    = \frac{\mathrm{d}\sigma_{\mathrm{partons}}(O - a_O {\mathcal P})}{\mathrm{d}O}, $
where the power correction ${\mathcal P}$ being universal for all event shapes, and $a_O$ being specific known constants~\cite{this}. 
As a result,
$
\langle O^1 \rangle_{\mathrm{hadrons}}
    = \langle O^1 \rangle_{\mathrm{partons}} + a_O \mathcal{P}\,,
$
with $\langle O^1 \rangle_{\mathrm{partons}}$  obtained as described in Sect.~\ref{sec:theory}.
The expression $\mathcal{P}$ depends on theoretically calculable constants, namely  
the so-called ``Milan factor'' 
and the value of effective coupling below the low fixed scale $\mu_I = 2\GeV$, $\alpha_0(\mu_I)$,
which is a non-perturbative parameter of the dispersive model and can be related to the 
effective soft coupling 
$\alpha_{S}^{CMW}$ (Catani-Webber-Marchesini scheme).
The relation between the strong coupling defined in the 
$\overline{\mathrm{MS}}$ scheme and the effective soft coupling 
$\alpha_{S}^{CMW}$ is scheme-dependent and complex at higher orders~\cite{Banfi:2018mcq,
Catani:2019rvy}. 
However, in one particular scheme, the relation between $\alpha_{S}$ 
and $\alpha_{S}^{CMW}$ has recently been computed up to $\mathcal{O}(\alpha_{S}^{4})$ 
accuracy~\cite{Catani:2019rvy} ($A^0$ scheme), which allow for an implementation of a consistent analytic 
model of hadronization corrections at order that matches the order of pQCD predictions.
For the details of the implementation see Ref.~\cite{this}.
For the qualitative comparison with the models based on the MCEG approach, the corrections obtained with the  $A^0$-scheme are 
transformed into multiplicative factors and presented in Fig.~\ref{fig:had}.

\section{Results and discussion}
\label{sec:results}
\begin{figure}[htbp]\centering
\includegraphics[width=\BROADFIGWIDTH]{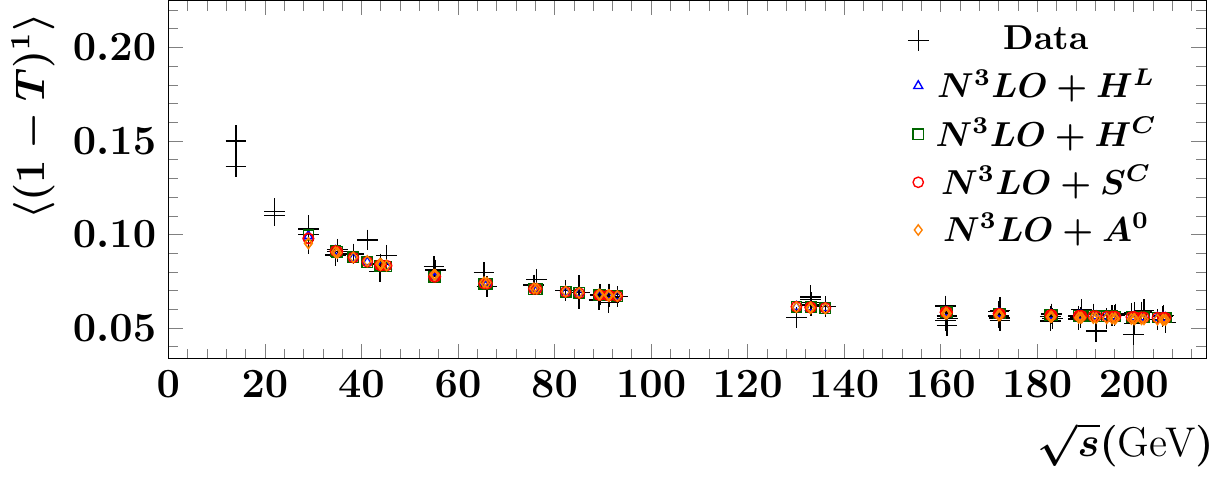}\includegraphics[width=\BROADFIGWIDTH]{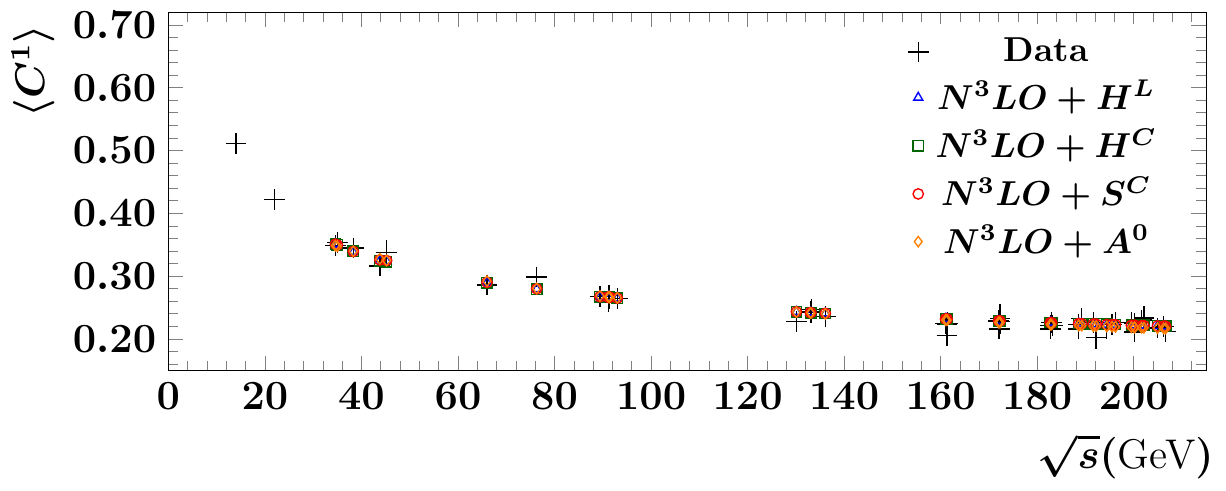}\\
\caption{Data and fits to the data obtained with different types of hadronization models. Figures from Ref.~\protect\cite{this}.}
\label{fig:result}
\end{figure}

The values of $\alpha_{S}(M_Z)$  were determined in the optimization procedures as described in Ref.~\cite{this}
using multiple data sets  from $e^{+}e^{-}$ collision experiments at LEP, PETRA, PEP and TRISTAN colliders. 
The multiple numerical results of the NNLO and N${}^3$LO fits are presented in Ref.~\cite{this},  
while the predictions of the N$^{3}$LO fits for individual energy 
points are shown in Fig.~\ref{fig:result}.
The NNLO results for $\alpha_{S}(M_Z)$ obtained with all MCEG and analytic 
hadronization models are in good agreement between the fits to $\langle (1-T)^1 \rangle$ 
and $\langle C^1 \rangle$, which can be viewed as a check of the  consistency of 
the $\alpha_{S}(M_{Z})$ extraction method at NNLO. Similarly to previous 
studies~\cite{Gehrmann:2009eh},
a discrepancy between the results obtained with the MCEGs hadronization models and 
the analytic hadronization models are seen.
While the parameters $\alpha_0(2\GeV)$  in different schemes do not represent the ``same'' quantity 
and cannot be 
compared compared,  their numerical values obtained in the fits are numerically very similar.

The obtained results with N${}^{3}$LO predictions  have similar patterns to the results obtained with 
the NNLO predictions.
 However, as expected, the obtained uncertainties are somehow larger than for the corresponding quantities 
 obtained with NNLO predictions.
The values $D^{\langle (1-T)^1 \rangle}$ and 
$D^{\langle C^1 \rangle}$  obtained with different hadronization models reasonably agree with each other, 
which could serve as an indirect evidence that the higher-order coefficients $D^{\langle O^n \rangle}$ can be extracted 
from the data even with higher  precision if more data would be available.
In the same time, the  differences between the  $\alpha_{S}(M_{Z})$ values  obtained with different types of hadronization 
models are preserved even at N${}^{3}$LO accuracy. 
This suggests that the discrepancy pattern has a 
fundamental origin and will hold in the future analyses even with more data and exact N${}^{3}$LO predictions available.

Hereby, improvement of the hadronization modeling and a better understanding of 
hadronization itself is more important for increasing the precision of 
$\alpha_{S}(M_{Z})$ extractions than the calculation of perturbative 
corrections beyond NNLO. 
\section{Conclusions}
\label{sec:conclusions}
We discussed the extraction of the $\alpha_{S}(M_{Z})$ from 
 available data on the averages of event shapes 
 $\langle (1-T)^1 \rangle$ and $\langle C^1 \rangle$ in N${}^3$LO  and NNLO  accuracy in pQCD using different types of hadronization models.
The results obtained using NNLO predictions and analytic hadronization 
corrections based on the dispersive model are consistent with the recent world 
average.

The method of extraction of $\alpha_{S}(M_{Z})$ in  
N${}^3$LO precision in pQCD  uses a combination of  NNLO predictions calculated from the first principles  
and estimations of the N${}^3$LO contributions from the data. 
The method produced results which are compatible with the current world average 
within the somewhat large uncertainties,
e.g.\  the result from the fits to the $\langle (1-T)^1 \rangle$ data  reads $\alpha_{S}(M_{Z})^{N^{3}LO+A^0}=\TTbMOMresultNNNLO$ .
The obtained precision can be increased with more high-quality data from future experiments. 

In the discussed analysis the hadronization corrections were derived from the 
 MCEGs models and analytic hadronization models extended to higher orders for the first time.
The results obtained with those two approaches imply that future analyses will be strongly affected by the
hadronization effects even if the  exact higher -order corrections will be available.

However, in the last decades the developments in the modeling of particle collision by MCEGs 
were driven by the need to model the processes at high energies of 
LEP, HERA and LHC colliders and therefore had limited impact on the description of 
phenomena at lower energies.  Similarly,  it is expected that rapid developments in the modeling of particle 
collisions at lower energies and understanding of hadronization can be expected only 
with the availability of new measurements in the corresponding (lower) energy ranges.
In the context of future $e^{+}e^{-}$ colliders it can be achieved with a 
program of measurements of the hadronic final state properties at $\sqrt{s}\approx 20-50\GeV$ performed 
with radiative events or in dedicated collider runs.

\section*{Acknowledgements and Funding information}
A.K. acknowledges financial support from the Premium Postdoctoral 
Fellowship program of the Hungarian Academy of Sciences. This work was 
supported by grant K 125105 of the National Research, Development and
Innovation Fund in Hungary.
{\tiny
\bibliography{MOM.bib}

\begin{thebibliography}{1}
\providecommand{\url}[1]{\texttt{#1}}
\providecommand{\urlprefix}{URL }
\expandafter\ifx\csname urlstyle\endcsname\relax
  \providecommand{\doi}[1]{doi:\discretionary{}{}{}#1}\else
  \providecommand{\doi}{doi:\discretionary{}{}{}\begingroup
  \urlstyle{rm}\Url}\fi
\providecommand{\eprint}[2][]{\url{#2}}

\bibitem{GehrmannDeRidder:2007hr}
{A.~Gehrmann-De Ridder et al.},
\newblock \emph{{{NNLO} corrections to event shapes in $e^{+} e^{-}$
  annihilation}},
\newblock JHEP \textbf{12}, 094 (2007),
\newblock \doi{10.1088/1126-6708/2007/12/094},
\newblock \eprint{0711.4711}.

\bibitem{Baikov:2012zn}
{P.A.~Baikov et al.},
\newblock \emph{{Adler Function, Sum Rules and Crewther Relation of Order
  {${\cal O}(\alpha_s^{4})$}: the Singlet Case}},
\newblock Phys. Lett. \textbf{B714}, 62 (2012),
\newblock \doi{10.1016/j.physletb.2012.06.052},
\newblock \eprint{1206.1288}.

\bibitem{DelDuca:2016csb}
{V.~Del Duca et al.},
\newblock \emph{{Three-jet production in electron-positron collisions at
  next-to-next-to-leading order accuracy}},
\newblock Phys. Rev. Lett. \textbf{117}(15), 152004 (2016),
\newblock \doi{10.1103/PhysRevLett.117.152004},
\newblock \eprint{1603.08927}.

\bibitem{this}
A.~Kardos, G.~Somogyi and A.~Verbytskyi,
\newblock \emph{{Determination of $\alpha _{S}$ beyond NNLO using event shape
  averages}},
\newblock Eur. Phys. J. C \textbf{81}(4), 292 (2021),
\newblock \doi{10.1140/epjc/s10052-021-08975-3},
\newblock \eprint{2009.00281}.

\bibitem{Dokshitzer:1995qm}
{Y.L.~Dokshitzer, G.~Marchesini and B.R.~Webber},
\newblock \emph{{Dispersive approach to power behaved contributions in QCD hard
  processes}},
\newblock Nucl. Phys. B \textbf{469}, 93 (1996),
\newblock \doi{10.1016/0550-3213(96)00155-1},
\newblock \eprint{hep-ph/9512336}.

\bibitem{Banfi:2018mcq}
{A.~Banfi, B.K.~El-Menoufi and P.F.~Monni},
\newblock \emph{{The Sudakov radiator for jet observables and the soft physical
  coupling}},
\newblock JHEP \textbf{01}, 083 (2019),
\newblock \doi{10.1007/JHEP01(2019)083},
\newblock \eprint{1807.11487}.

\bibitem{Catani:2019rvy}
{S.~Catani, D.~De Florian and M.~Grazzini},
\newblock \emph{{Soft-gluon effective coupling and cusp anomalous dimension}},
\newblock Eur. Phys. J. C \textbf{79}(8), 685 (2019),
\newblock \doi{10.1140/epjc/s10052-019-7174-9},
\newblock \eprint{1904.10365}.

\bibitem{Gehrmann:2009eh}
{T.~Gehrmann, M.~Jaquier and G.~Luisoni},
\newblock \emph{{Hadronization effects in event shape moments}},
\newblock Eur. Phys. J. \textbf{C67}, 57 (2010),
\newblock \doi{10.1140/epjc/s10052-010-1288-4},
\newblock \eprint{0911.2422}.

\end{thebibliography}
}
\nolinenumbers
\end{document}